\documentclass[a4paper,showkeys,floatfix,aps,prl,reprint,groupedaddress]{revtex4-1}
\usepackage{amsmath,amssymb}
\usepackage{graphics,graphicx}
\usepackage{dcolumn,bm}
\usepackage{psfrag}
\usepackage{soul}
\usepackage{hyperref}
\usepackage{float}
\usepackage{paralist}
\usepackage{array}
\usepackage[table]{xcolor}
\usepackage{makecell}
\newcolumntype{P}[1]{>{\centering\arraybackslash}p{#1}}
\hypersetup{colorlinks=true, citecolor=blue, urlcolor=blue, linkcolor=blue}
\usepackage{colortbl}

\begin{document}

\title{Critical crack-length during fracture}

\author{Viswakannan R. K.}
\email{p20230074@hyderabad.bits-pilani.ac.in}
\author{Subhadeep Roy}
\email{subhadeep.r@hyderabad.bits-pilani.ac.in}
\affiliation{Department of Physics, Birla Institute of Technology and Science Pilani, Hyderabad Campus, Secunderabad 500078, Telangana, India.}

\date{\today}

\begin{abstract}
     
Through controlled numerical simulations in a one dimensional fiber bundle model with local stress concentration, we established an inverse correlation between the strength of the material and the cracks which grow inside it - both the maximum crack and the one that set in instability within the system, defined to be the critical crack. Through Pearson correlation function as well as probabilistic study of individual configurations, we found that the maximum and the critical crack often differ from each other unless the disorder strength is extremely low. A phase diagram on the plane of disorder vs system size demarcates between the regions where the largest crack is the most vulnerable one and where they differ from each other but still shows moderate correlation.
\end{abstract}

\pacs{}

\maketitle


Disorder plays a crucial role during the failure of heterogeneous media, a topic that has been studied extensively in last few decades \cite{arcangelis78,duxbury86,herrmann90,pradhan10}. The existence of defects like micro-cracks and the interactions among them makes the process of crack propagation more complicated than Griffith's theory, which was suggested by A. A. Griffith in 1921 and reported the critical stress of a homogeneous media with a pre-existing crack of length $l$ to vary as $1/\sqrt{l}$. For heterogeneous media, on the other hand, Griffith's criterion produces significant error in determining the critical stress or surface energy due to the resistance in the form of an energy barrier that ultimately arrests a propagating crack, widely known as the lattice trapping or intrinsic crack resistance \cite{bernstein03,mattoni05,perez00}. Two length scales are observed to emerge as a result of such lattice trapping: a small length scale related to the dissipation of energy near the crack tip and the a large length scale associated with the elastic deformation around the tip \cite{bernstein03,mattoni05,curtin00,long21}. Even for perfectly brittle materials, one needs to take into account the discrete atomistic nature of the interactions and make modifications in Griffith's theory \cite{broberg99}. A correlation between crack length and nominal stress is very important for damage control as it can provide necessary information regarding an upcoming catastrophic failure. Application of such prediction ranges from laboratory experiments to large scale building blocks and even geological scales like seismic events \cite{wu18}. The micro and meso-scale heterogeneity not only affects the nominal stress but also the course of failure by introducing local breaking events, known as avalanches, producing crackling noises which can be captured in an acoustic emission (AE) experiment \cite{baro13}. The AE process includes the translation of the crackling noises (during a crack propagation) into bursts and subsequently emitted energies, known as the acoustic signals or precursors. Such precursors become more populated in an accelerating manner \cite{roy23} as one approaches the global failure indicated by the increasing rate of deformation \cite{petri94} through gradual accumulation of damages \cite{ramos13}. 

A classic work discussing the propagation of smaller crack vs larger crack has been published by Paris in 1963 where the cracks bigger than a critical length were observed to follow the Paris-law and not the others \cite{paris63}. This is followed by works of Kitagawa and Takahashi \cite{kitagawa76} where a scale-free decay of critical stress for larger crack lengths, similar to Griffith's law, was observed with a crack-length independent critical stress for smaller cracks \cite{haddad80}. Later, Taylor \cite{taylor81} showed the existence of two length scales in the work by Kitagawa and Takahashi. A similar existence of length scales are observed \cite{roy22} numerically as well by one of the author of the present paper in a statistical disorder system, the fiber bundle model \cite{book_fbm}, acted by a tensile force in presence of local stress concentration.   

\begin{figure}[ht]
\centering
\includegraphics[width=9.0cm, keepaspectratio]{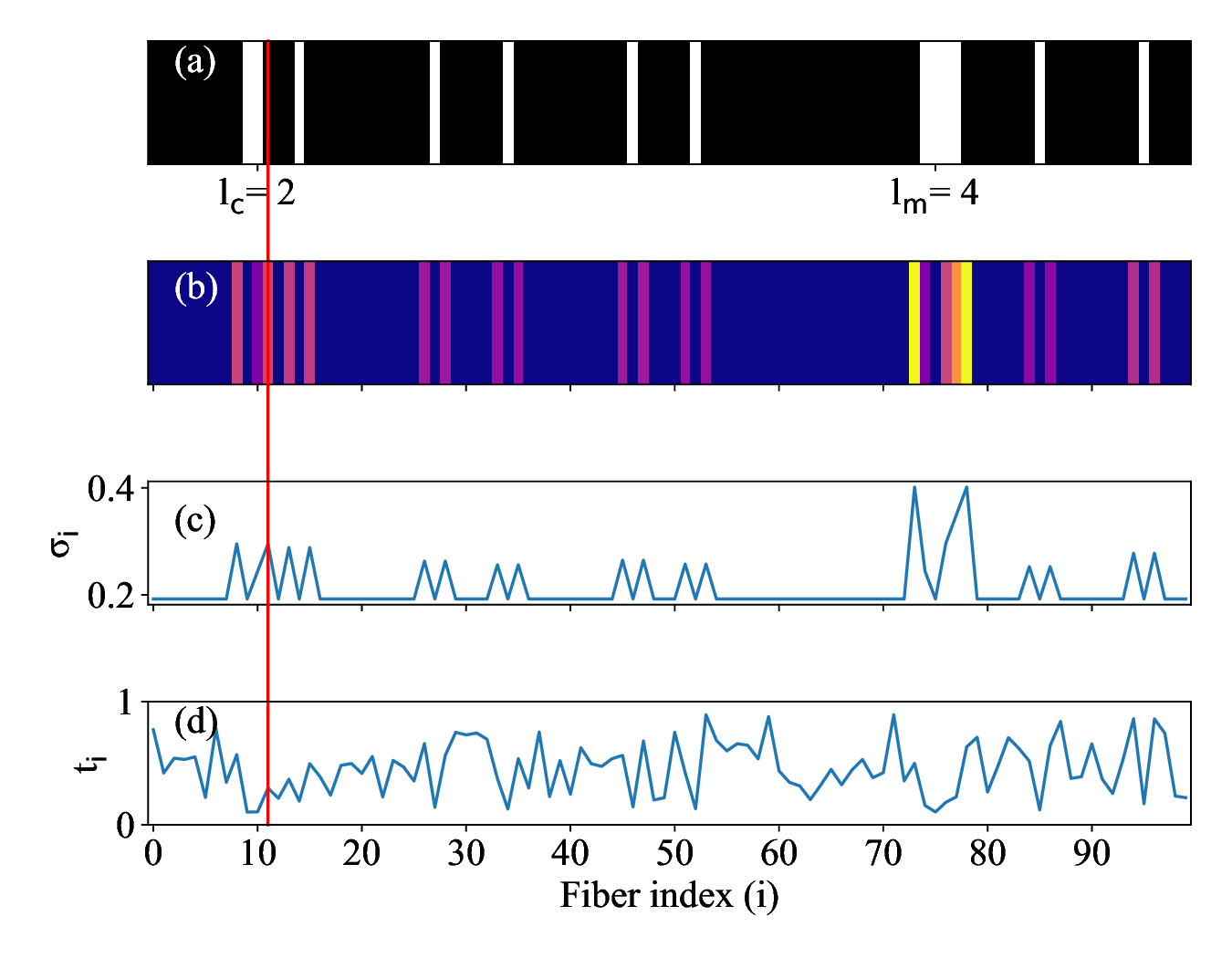}
\caption{(a) The status of a fiber, broken or intact, is represented by white and black color respectively against the fiber index. (b) The heat map for the local stress profile - blue color stands for lower local stress while the yellow for relatively higher stress. (c)-(d) Local stress and threshold values as a function of fiber indices. The vertical line corresponds to the fiber that breaks and set instability within the system. We kept $\delta=0.4$ and $L=10^2$.}
\label{fig1}
\end{figure}

This paper focuses on the numerical study of a disordered system, the fiber bundle model \cite{book_fbm}, as a prototype for failure dynamics in heterogeneous media. The model consists of $L$ vertical fibers attached between two horizontal soft clamps pulled apart by a stress $F$, exerting a stress $\sigma=F/L$ on each fiber. Each fiber has an unique threshold stress chosen from a uniform distribution spanning from $0.5-\delta$ to $0.5+\delta$, $\delta$ is the strength of disorder and $0.5$ is the mean of the distribution. Being stretched beyond this threshold, a fiber breaks irreversibly and the stress carried by the broken fiber is redistributed among its neighbouring fibers as per the following rule: 
\begin{align}
\sigma_r \rightarrow \sigma_r + \displaystyle\frac{\sigma_bd_l}{d_r+d_l} \nonumber \\
\sigma_l \rightarrow \sigma_l + \displaystyle\frac{\sigma_bd_r}{d_r+d_l}
\end{align}
Here, $\sigma_b$ is the stress of the broken fiber, $\sigma_r$ and $\sigma_l$ are respectively the local stress of the right and left nearest neighbor. The stress distribution is made distance dependent to eliminate any memory/history dependence in the process. We have used the minimum image convention to include the boundary effect due to the periodic boundary condition and calculate the actual distances $d_l$ and $d_r$. Such a local stress redistribution is a result of the soft membrane which is supporting the fibers and mimics the nature of stress localization in an elastic media \cite{hui13}. The redistribution can induce further rupture events, starting an avalanche, due to the local stress enhancement until the next threshold is beyond the redistributed stress. The external force, at this moment, is increased in a quasi-static manner to break the next weakest fiber and the model evolves through a number of stress increment and avalanches until all fibers break suggesting the global failure. The final value of externally applied stress just before global failure is the critical stress or strength of the bundle. The critical crack-length is the size of the crack in the last stable configuration (just before global failure) which propagates and set in instability starting the final avalanche and hence breaking the rest of the bundle. The critical crack is one among a number of micro-cracks and statistically the most vulnerable one. This does not guarantee that this critical crack will be maximum in length. We will discuss this in details next.    

\begin{figure}[ht]
\centering
\includegraphics[width=8.0cm, keepaspectratio]{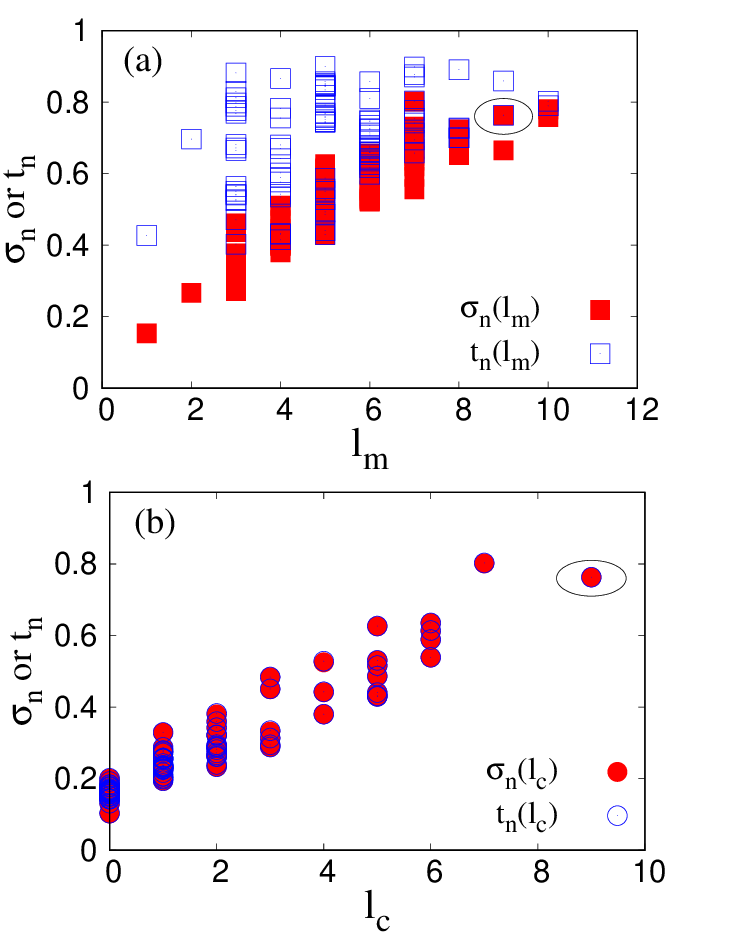}
\caption{Comparison between local stress and threshold values at the notches for the (a) maximum and (b) critical crack. The figures show 100 different configurations for $L=10^3$ and $\delta=0.4$. The ellipse shows a particular configuration for which the critical and maximum cracks are the same one.}
\label{fig1a}
\end{figure}

Here we will discuss whether there exists a relationship between the nominal stress, the stress at which the heterogeneous system breaks, and the critical crack that initiates the final avalanche. Let's call the former $\sigma_c$ and latter $l_c$. The maximum crack length, on the other hand is denoted by $l_m$ and in principle it can be different from $l_c$. Through detailed numerical simulation in a 1d fiber bundle model of size $L$ and disorder strength $\delta$, we have explored the nature of critical crack length $l_c$ and maximum crack length $l_m$ and how they respond as we change either of the above two parameters, $L$ or $\delta$. To be specific, whether there is a scope of future failure prediction associated with it. In gist, we pose a two-fold question here: 
\begin{compactitem}
\item Does the length of the most vulnerable crack contain any information regarding the nominal stress and vice versa? 
\item Is the most vulnerable crack always the largest one or does it depends on the material properties?    
\end{compactitem}
A boundary on the $L-\delta$ plane shows the region where the maximum and the vulnerable cracks are interlinked and where they are mutually exclusive. 

Figure \ref{fig1} shows the micro-cracks which are developed within the 1d chain just before the global failure. We chose $\delta=0.4$ and $L=10^2$. The smaller system size is adopted to make the micro-cracks more visible. Later we have used higher system sizes for the numerical simulation. This diagram will be used to establish the fact that due to local stress concentration, the local stress to threshold difference at the notches of the micro-cracks is mutually exclusive from the size of the cracks itself. Figure \ref{fig1}(a) shows the broken and the intact fibers by white and black color respectively. The biggest white patch stands for the maximum crack $l_m$, which is of length 4 (fiber indices 74 to 77) for the present configuration. The critical crack, $l_c$, for the same configuration is of length 2 (fiber indices 9 \& 10) on the other hand. The local stress profile for the same is shown in figure \ref{fig1}(b) - lower to higher local stress as we go from blue to yellow colors. The red vertical line represents the fiber (fiber index 11) that breaks and set in instability. The fact that, in-spite of having higher notch stresses ($=0.401639$) around the maximum crack, it does not propagate, is really counter intuitive and monitoring the maximum crack can be highly misleading for failure prediction and damage control. The critical crack propagates in-spite of having a lower notch stresses ($=0.295405$). 

\begin{figure}[ht]
\centering
\includegraphics[width=8.0cm, keepaspectratio]{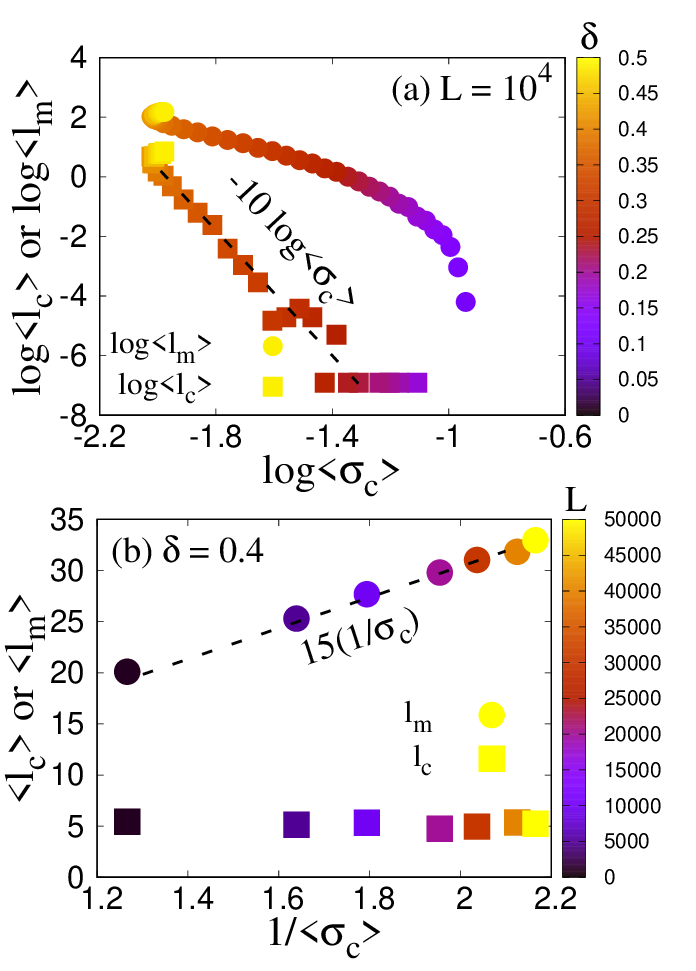}
\caption{(a) Decreasing trend of $\langle l_c \rangle$ and $\langle l_m \rangle$ with critical stress $\langle \sigma_c \rangle$ for $L=10^4$ and varying disorder. While $\langle l_m \rangle$ does not show a particular trend, $\langle l_c \rangle$ shows the following behavior: $\langle l_c \rangle \sim \langle \sigma_c \rangle^{-\eta}$, where $\eta \approx 10$. (b) For $\delta=0.4$ and varying system sizes, $\langle l_c \rangle$ remains constant w.r.t $\langle \sigma_c \rangle$ while $\langle l_m \rangle \sim \langle \sigma_c \rangle^{-1}$.}
\label{fig2}
\end{figure}

This is due to the interplay between the local stress and threshold values of individual fibers. Figure \ref{fig1}(d) shows that the threshold values at the notches of the maximum and critical crack are $0.632846$ and $0.298248$ respectively. Due to this, the local stress to threshold difference around the maximum crack ($0.231207$) is much higher than that of the critical crack ($0.002843$), making the critical crack much more prone to propagation. Such a stress to threshold comparison is more evident from figure \ref{fig1a} where the notch stresses $\sigma_n$ and threshold values $t_n$ are plotted against $l_m$ (see figure \ref{fig1a}a) and $l_c$ (see figure \ref{fig1a}b) for 100 different configurations. It is clear that for most of the configurations, $t_n>\sigma_n$ for $l_m$ ranging between 0 to 12 with higher values more populated. On the other hand, we observe an overlap between $t_n$ and $\sigma_n$ independent of the values of $l_c$, which spans from 0 to 10 and populated towards the lower values. We have circled a particular configuration where the critical and maximum crack is the same one, though such configurations are very rare to find at a moderate disorder ($\delta=0.4$).         

\begin{figure}[t]
\centering
\includegraphics[width=8.0cm, keepaspectratio]{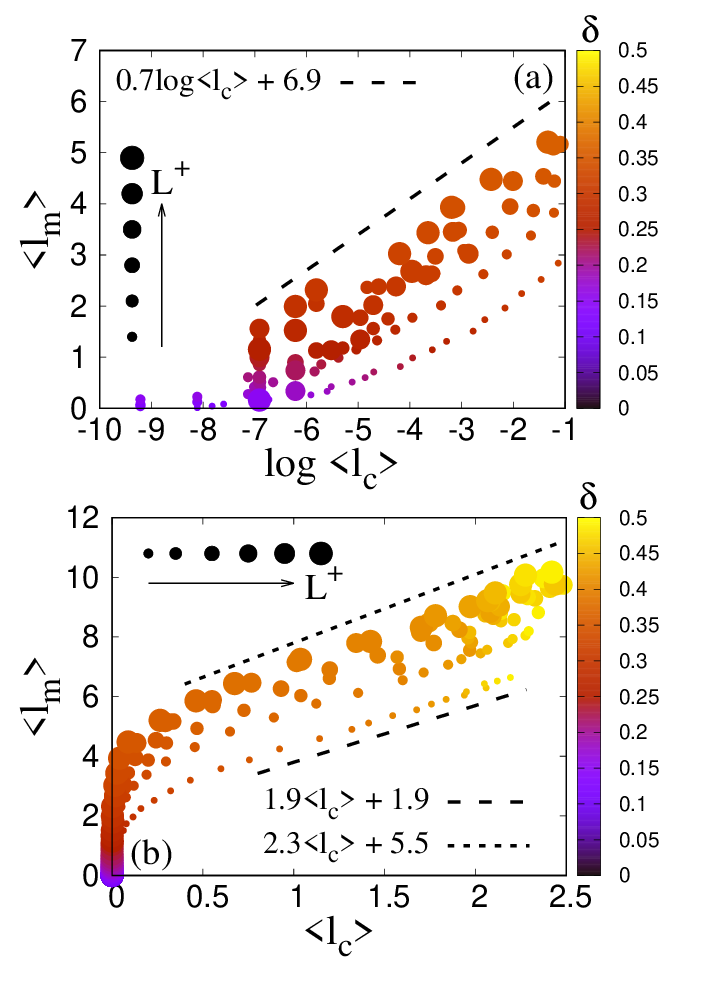}
\caption{Correlation between average critical crack $\langle l_c \rangle$ and average maximum crack $\langle l_m \rangle$ for different system sizes and disorder strength. We observe different behaviors at small and large length scales. (a) For small scale: $\langle l_m \rangle \sim \log \langle l_c \rangle$. (b) For large scale: $\langle l_m \rangle$ grows linearly with $\langle l_c \rangle$. The color gradient is on the disorder strength while point size for the size of the system which varies from $10^3$ (smallest circles) to $5\times10^4$ (largest circles).}
\label{fig3}
\end{figure}

\begin{figure*}[t]
\centering
\includegraphics[width=14.0cm, keepaspectratio]{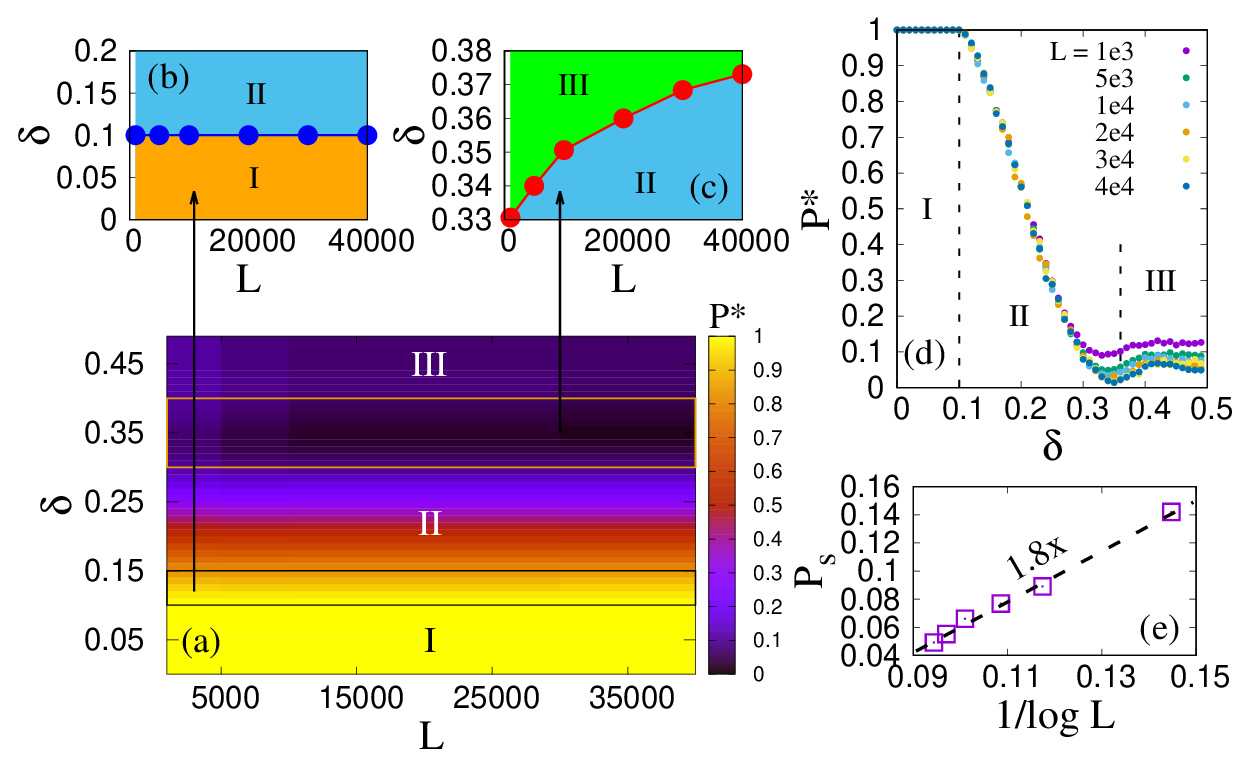}
\caption{(a) Probability $P^{\ast}$ that $l_c=l_m$ on the $\delta - L$ plane. (b)-(c) The I-II and II-III phase boundaries are zoomed in for better visibility (d) $P^{\ast}$ vs $\delta$ for system sizes ranging between $10^3$ and $4\times 10^4$. (a)-(d) shows 3 distinct regions: (I) $P^{\ast}=1$ independent of both $\delta$ and $L$; (II) $P^{\ast}$ is a decreasing function of $\delta$ without a L dependence; (III) $P^{\ast}$ is a constant independent of $\delta$ with a slight $L$-dependence. (e) System size effect of $P_s$, the saturation of $P^{\ast}$ at high disorder: $P_s = \displaystyle\frac{1.8}{\log L}-0.07$.}
\label{fig4}
\end{figure*}

Figure \ref{fig2} shows the correlation between the average critical stress $\langle \sigma_c \rangle$ and cracks, both critical and maximum, developed within the 1d chain. We keep the system size constant at $L=10^4$ and vary the disorder strength $\delta$ from 0 to 0.5. The critical stress is observed to decrease with increasing length of critical crack or the maximum crack. Such reduction of critical stress is observed earlier in real systems like mild steels under s periodic load \cite{taylor84}. In the fiber bundle model, the critical stress was observed to fall in a same scale-free manner with the length of the pre-existing crack \cite{roy22}. The present scenario is much different as we are not starting with a pre-existing crack hence the process includes both crack initiation and propagation through the dynamics of the model only.  

Figure \ref{fig2}(a) registered the critical crack $\langle l_c \rangle$ and the maximum crack $\langle l_m \rangle$ as a function of $\langle \sigma_c \rangle$ when we tune the disorder strength $\delta$ keeping the size of the bundle fixed at $L=10^4$. While both $\langle l_m \rangle$ and $\langle l_c \rangle$ decreases with $\langle \sigma_c \rangle$ as $\delta$ decreases, $\langle l_c \rangle$ only follows the trend below as move towards lower disorder strength: 
\begin{align}
\langle l_c \rangle \sim \langle \sigma_c \rangle^{-\eta}
\end{align}
with $\eta \approx 10$. Both $\langle l_c \rangle$, and $\langle l_m \rangle$ decreases with decreasing $\delta$ as the failure process becomes more and more brittle-like where the bundle breaks abruptly at a relatively higher stress. On the other hand, if we keep the disorder strength constant ($=0.5$) and vary the system size between $10^3$ and $5 \times 10^4$, $\langle l_c \rangle$ remains constant independent of the critical stress. At the same time, $\langle l_m \rangle$ changes as follows: 
\begin{align}
\langle l_m \rangle \sim \langle \sigma_c \rangle^{-1}
\end{align}
The maximum crack grows with the system size as a higher system size will allow the thresholds come closer to each other as per the weakest link of chain theory \cite{ray85} and at the same time increases the density of the strong as well as weak fibers. The weakest link of chain will allow a crack to propagate and the higher density of stronger fibers increases the chance of a large crack getting arrested as a result of higher threshold value relative to notch stress. The critical crack-length remains same independent of the critical stress as the critical crack becomes unstable since it did not encountered a high enough threshold but still experiences the weakest-link-of-chain event. As system size increases the critical stress decreases but a crack-length of almost a similar size is proven to be enough to create instability at the notches.

Next we turn to understand how the two crack lengths, critical and maximum, are correlated to each other. For this, we have adopted three different ways: (i) comparing the average values (over $10^4$ realizations) of $l_m$ and $l_c$ denoted by $\langle l_m \rangle$ and $\langle l_c \rangle$ respectively, (ii) probabilistic approach for individual realizations and the $l_m$ and $l_c$ we find in each realization, and (iii) through a Pearson correlation function. We will discuss these three approaches one by one next.

Figure \ref{fig3}(a) and (b) shows the correlation between $\langle l_c \rangle$ and $\langle l_m \rangle$ both at a smaller as well as at a larger scale. The color gradient represents the disorder strength and the size of the points stand for the system size - larger the points higher the size of the bundle. We have used system sizes ranging in between $10^3$ (smallest circles in figure \ref{fig3}) and $5\times10^4$ (largest circles in figure \ref{fig3}). For small lengths of both critical and maximum cracks, we observe the following logarithmic dependence at a sufficiently larger system size
\begin{align}
\langle l_m \rangle = a\log \langle l_c \rangle + b
\end{align}
where $a=0.7$ and $b=6.9$. At a larger scale, on the other hand, such logarithmic dependence vanishes and instead we observe a linear behavior between $\langle l_m \rangle$ and $\langle l_c \rangle$
\begin{align}
\langle l_m \rangle = c\langle l_c \rangle + d
\end{align}
where $c$ increases from 1.9 to 2.3 as system size is increased from $10^3$ to $4\times 10^4$. For the same increment in $L$, $d$ increases from 1.9 to 5.5. 

Next, we study the correlation between $l_c$ and $l_m$ from a probabilistic approach in order to draw a phase diagram on the $L-\delta$ plane. This is represented in figure \ref{fig4}. We define $P^{\ast}$ as the probability that the critical crack and the maximum crack are the same. In other word this is the probability that the instability within the system is created by the maximum crack. Fig \ref{fig4}(a) shows the heat map of $P^{\ast}$ on the $L-\delta$ plane where \ref{fig4}(d) explicitly shows how $P^{\ast}$ varies with $\delta$ for system sizes ranging between $10^3$ and $4 \times 10^4$. Specifically, we observe the following three distinct regions. 
\begin{compactitem}
\item Region I: In this region, $P^{\ast}=1$ making $l_m$ equals to $l_c$ for each and every configuration. This is an extremely brittle region where the first fiber initiate global failure each and every time. This boundary between I and II remains constant at 0.1 independent of the size $L$ of the bundle (see figure \ref{fig4}b).
\item Region II: The probability of $l_m$ being equal to $l_c$ is less than 1 here and a decreasing function of disorder strength, but do not respond to the change in system sizes. The boundary, unlike the boundary between I and II, is dependent on system size - for higher $L$, we have to go to a higher disorder strength to enter region III. This makes sense since increasing $L$ makes the failure process more abrupt and the disorder strength has to be increases to compensate for that.  
\item Region III: In this region, $P^{\ast}$ reaches its minimum and saturates afterwards at a constant value $P_s$ independent of the disorder strength. Contrary to region II, here $P^{\ast}$ is a decreasing function of system size. Figure \ref{fig4}(e) shows the following scaling as we approach the thermodynamic limit:
\begin{align}
P_s = \displaystyle\frac{1.8}{\log L}-0.07
\end{align}
The constant value being closer to zero suggest that $P_s$ becomes zero as we approach $L \rightarrow \infty$. This means in the thermodynamic limit, if we are at region III, we will not be able to find any configuration where the instability is set on by the maximum crack length.
\end{compactitem} 

The final thrust to the correlation study will be calculating the Pearson correlation function directly as we tune both $\delta$ and $L$ over $10^3$ realizations for each set of ($\delta$, $L$). The correlation function defined below gives us an even better idea of how strong or weak the correlation is in between critical and maximum crack for individual realizations and not after taking the average. This way it can be more relatable to the experiments since a single experiment can be considered as a single realization in our simulation. The correlation function has the following form:
\begin{align}
c_p = \displaystyle\frac{\displaystyle\sum (x_i-\bar{x})(y_i-\bar{y})}{\left[\displaystyle\sum (x_i-\bar{x})^2 \displaystyle\sum (y_i-\bar{y})^2\right]^{1/2}}  
\end{align}
where $x_i$ and $y_i$ values represent $l_c$ and $l_m$ for individual configurations while $\bar{x}$ and $\bar{y}$ are their average values $\langle l_c \rangle$ and $\langle l_m \rangle$. 

\begin{figure}[ht]
\centering
\includegraphics[width=8.0cm, keepaspectratio]{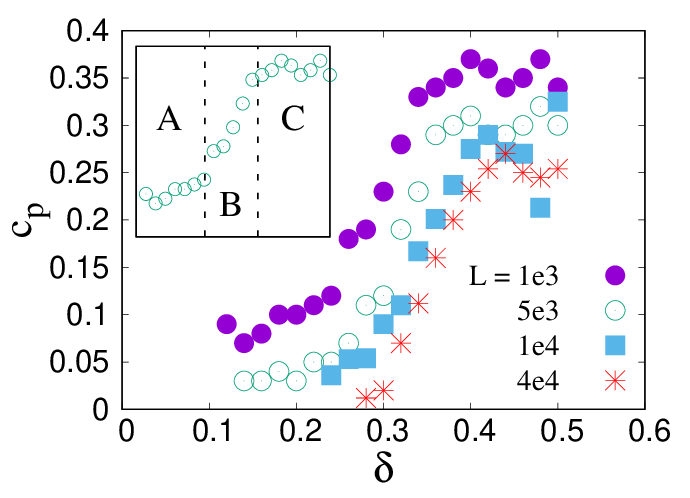}
\caption{Pearson correlation coefficient $c_p$ as a function of disorder strength $\delta$ for system sizes ranging from $10^3$ to $4\times10^4$. The inset shows the result for $L=5\times10^3$ with three distinct regions: (A) constant low correlation, (B) correlation increasing with disorder, and (C) constant high correlation.}
\label{fig5}
\end{figure}

\begin{figure}[ht]
\centering
\includegraphics[width=8.0cm, keepaspectratio]{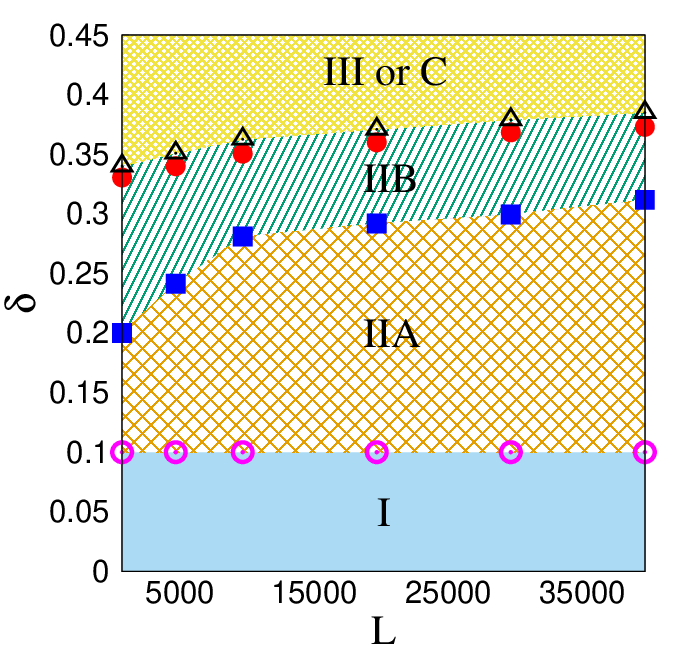}
\caption{$L-\delta$ plane showing different regions depending of the interplay between $l_c$ and $l_m$. \\ Region I: $l_c=l_m$ with a probability 1. $c_p$ is undefined here as there is no fluctuation in the realizations. \\ Region IIA: combination of region II and region A, making $P^{\ast}$ a decreasing function of $\delta$ but $c_p$ constant at a low value. \\ Region IIB: combination of region II and region B, making $P^{\ast}$ a decreasing function but $c_p$ an increasing function of $\delta$. \\ Region III or C: III and C are equivalent giving $P^{\ast}$ constant at a low value or $c_p$ constant at a high value. The boundary of IIB and III (or C) is drawn from both the study of $P^{\ast}$ (red solid circles) as well as $c_p$ (black hollow triangle).}
\label{fig6}
\end{figure}

The results are shown in figure \ref{fig5}. The Pearson coefficient $c_p$ is an increasing function of $\delta$. This makes sense since for high $\delta$, the failure process happens through a number of avalanches giving us enough information to correlation between $l_c$ and $l_m$. At a low $\delta$, on the other hand, the failure process is much more abrupt (like brittle material) and there are less chances for such prediction. Because of the same reason, with increasing system size, as the failure process becomes more abrupt, $c_p$ decreases. In the inset of the same figure, we divided the whole region for $\delta$ into the following three parts: 
\begin{compactitem}
\item Region $A$: $c_p$ is almost constant at a low value independent of $\delta$. $l_c$ and $l_m$ are loosely correlated in this region. 
\item Region $B$: $c_p$ increases with $\delta$.
\item Region $C$: $c_p$ remains constant but at a higher value closer to 0.4. The correlation between $l_c$ and $l_m$ is decent here. 
\end{compactitem}
In figure \ref{fig6}, we have extensively discussed how the region $C$ is equivalent to the region III in figure \ref{fig4}. Also, the region II in figure \ref{fig4} can be divided between $A$ and $B$ if we consider both the nature of probability $P^{\ast}$ and correlation function $c_p$.

Figure \ref{fig6} shows a detailed phase diagram on the disorder vs system size plane by considering the contributions of both the probability $P^{\ast}$ as well as the correlation function. This splits the whole $L-\delta$ plane in four regions: I, IIA, IIB and III (or C). We have discussed all four regions below with extreme detail. We have also provided the description in a tabular form below.


\begin{table}[h!]
\centering
\begin{tabular}{| c | c | c | c |} 
 \hline
\cellcolor{gray!30} Region I & \cellcolor{gray!30} Region IIA & \cellcolor{gray!30} Region IIB & \cellcolor{gray!30} Region III/C \\ 
 \hline\hline
 \makecell{\cellcolor{blue!30} \\ \cellcolor{blue!30}\cellcolor{blue!30} $P^{\ast}$ constant \\ \cellcolor{blue!30} at 1 \\ \cellcolor{pink!70} \\ \cellcolor{pink!70} $c_p$ can not \\ \cellcolor{pink!70} be defined} & \makecell{\cellcolor{blue!30} \\ \cellcolor{blue!30} $P^{\ast}$ decreases \\ \cellcolor{blue!30} with $\delta$ \\ \cellcolor{pink!70} $c_p$ remains at \\ \cellcolor{pink!70} a constant \\ \cellcolor{pink!70} low value} & \makecell{\cellcolor{blue!30} \\ \cellcolor{blue!30} $P^{\ast}$ decreases \\ \cellcolor{blue!30} with $\delta$ \\ \cellcolor{pink!70} \\ \cellcolor{pink!70} $c_p$ increases \\ \cellcolor{pink!70} with $\delta$} & \makecell{\cellcolor{blue!30} $P^{\ast}$ remains at \\ \cellcolor{blue!30} a constant \\ \cellcolor{blue!30} low value \\ \cellcolor{pink!70} $c_p$ remains at \\ \cellcolor{pink!70} a constant \\ \cellcolor{pink!70} high value} \\ \hline 
\end{tabular}
\caption{Table showing charectaristic behaviour of the regions: I, IIA, IIB and III (or C).}
\label{table:1}
\end{table}

We observe {\bf region I} when the disorder strength is extremely low. In this region the maximum stress always set in the instability. Not only that $l_m=0$ here, which means the instability sets in from any random point in the bundle and not from a crack tip. This is due to the low disorder which makes the failure process extremely fast and the model reaches the global failure even before the local stress concentration acts in. At a moderate disorder we find {\bf region II} which is again divided in two parts {\bf IIA} and {\bf IIB}. In {\bf IIA}, with increasing $\delta$, $P^{\ast}$ decreases making $l_m$ and $l_c$ more and more mutually exclusive. At the same time, $l_m$ and $l_c$ are loosely correlated here giving a low value of $c_p$ independent of $\delta$. In {\bf IIB}, $P^{\ast}$ shows the same behavior as {\bf IIA} but $c_p$ gradually increases here with $\delta$. Finally, in {\bf III}, $P^{\ast}$ saturates at a low value independent of $\delta$. Here almost for no realization the maximum crack is responsible for the instability and $l_c$ is most of the time different than $l_m$. At the same time, $c_p\approx0.4$ making $l_c$ and $l_m$ more than moderately correlated in this region. We want to stress the fact here that ${\bf III}$ and $C$ are same region on $L-\delta$ plane. This is evident from the boundary drawn between {\bf IIB} and {\bf III} (or {\bf C}) from the study of $P^{\ast}$ (red solid circles) as well as $c_p$ (black hollow triangle) and they almost fall on each other.       

\bigskip
The present study deals with the fact that during the failure process of a disordered system, it is not sufficient to monitor the largest crack in order to predict the instability in the system as often it might not be the vulnerable one and the instability in the system can be initiated from some other part, making failure prediction and damage control much more tricky than it already is. For proper failure prediction, both the knowledge of maximum and critical crack (the most vulnerable one) will be required simultaneously. In the present paper, we have numerically studied a fiber bundle model in one dimension with a varying disorder strength and system size. An inverse correlation is observed between the strength of the disordered media and two crack lengths which are prominent through during the failure process - critical crack length $l_c$ and the maximum crack length $l_m$. At the same time, the average $\langle l_m \rangle$ maximum crack and $\langle l_c \rangle$ of critical crack are correlated with each other linearly at a larger length scale and in a logarithmic way for a shorter length scale. Such a correlation between $\langle l_c \rangle$ and $\langle l_m \rangle$ on the $L-\delta$ plane, shows three distinct region. For low disorder, where the failure process is extremely abrupt, we get $l_c=l_m$ with unit probability and with an undefined correlation. 
With increasing disorder strength, $P^{\ast}$ decreases and $c_p$, the correlation between critical and maximum crack, increases. In this limit, the maximum and the critical crack becomes mutually exclusive and the chance that the final trigger comes from the maximum crack decreases. At the same time, $l_m$ and $l_c$ becomes moderately correlated ($c_p\approx0.4$ at higher disorder strength) and we can extract the information about one crack length from the other one with higher accuracy. The future direction to this work can be a controlled laboratory experiment monitoring both critical and maximum crack and correlation between them and use them for real life failure prediction and damage control.   

\bigskip

Viswakannan R. K. thanks Birla Institute of Technology and Science for the support provided during the work.


\end{document}